\documentstyle[12pt]{article}
\def\beq{\begin{equation}}
\def\eeq{\end{equation}}
\def\bea{\begin{eqnarray}}
\def\eea{\end{eqnarray}}

\def\ba{\begin{array}}
\def\ea{\end{array}}
\def\bce{\begin{center}}
\def\ece{\end{center}}

\begin{document}
\begin{titlepage}
\rightline{SNUTP-98-045}
\rightline{APCTP-98-018}
\rightline{UM-TG-209}
\rightline{hep-th/9806041}
\rightline{revised, Nov., 1998}
\def\today{\ifcase\month\or
January\or February\or March\or April\or May\or June\or
July\or August\or September\or October\or November\or December\fi,
\number\year}
\vskip 1cm
\centerline{\Large \bf  Branes, Orbifolds and the Three 
Dimensional}
\vskip .3cm  
\centerline{\Large \bf ${\cal N} = 2$ SCFT in the Large $N$ limit}
\vskip 1cm
\centerline{\sc Changhyun Ahn$^{a}$, 
Kyungho Oh$^{b}$ and Radu Tatar$^{c}$}
\vskip 1cm
\centerline{{\it $^a$ Center for Theoretical Physics, 
Seoul National University,
Seoul 151-742, Korea}}
\centerline{{\tt chahn@spin.snu.ac.kr}}
\centerline{{ \it $^b$ Dept. of Mathematics, University of Missouri-St. Louis,
St. Louis, MO 63121, USA }and}
\centerline{{ \it APCTP 207-43 Cheongyangri-dong, Dongdaemun-gu, Seoul 
130-012 Korea}}
\centerline{{\tt oh@arch.umsl.edu}}
\centerline{{\it $^c$ Dept. of Physics, University of Miami,
Coral Gables, FL 33146, USA} and}
\centerline{{\it Institute of Isotopic and Molecular Technology, 
3400 Cluj-Napoca, P.O. Box 700, Romania}}
\centerline{{\tt tatar@phyvax.ir.miami.edu}}
\vskip 2cm
\centerline{\sc Abstract}
\vskip 0.2in
We study the correspondence between the large $N$ limit of ${\cal N} = 2$ 
three dimensional superconformal
field theories and M theory on 
orbifolds of $AdS_4 \times {\bf S^7 }$. 
We identify the brane configuration
which gives ${\bf C^3/Z_3}$ as a background for the M theory
as a Brane Box Model or a $(p, q)$ web model.  
By taking the orbifold
projection on the known spectrum of Kaluza-Klein harmonics of supergravity, 
we obtain information about the chiral primary operators of the
orbifold singularities. 
\vskip 1in
\end{titlepage}
\newpage
\setcounter{equation}{0}

\section{Introduction}
\setcounter{equation}{0}

It has been  proposed in \cite{mal}  that
the large $N$ limit of super conformal field theories (SCFT) 
can be described by taking the
supergravity limit of the superstring compactified on anti-de Sitter 
(AdS) space.
The correlation functions of SCFT having 
the AdS boundary as its spacetime can be 
obtained from the dependence
of the supergravity action on the asymptotic behavior of fields at the
boundary \cite{fk,polyakov,witten}.
This way one can get the scaling dimensions of operators of SCFT from the 
masses of particles in string/M theory. In particular, 
${\cal N}=4$ super $SU(N)$ Yang-Mills theory in 4 dimensions is described by
type IIB string theory on $AdS_5 \times {\bf S^5}$. The gauge group can be
replaced by $SO(N)/Sp(N)$ \cite{witten1} by taking
appropriate orientifold operations \cite{aoy,kaku,fs}.

There are ${\cal N} =2, 1, 0$ superconformal models in 
4 dimensions
which have supergravity description when this is compactificatified 
on orbifolds of $AdS_5 \times {\bf S^5}$
\cite{kachru,vafa}. This proposed duality has been tested by studying
the Kaluza-Klein (KK) states of supergravity theory on the orbifolds of
$AdS_5 \times {\bf S^5}$ and by comparing them with the chiral 
primary operators
of the SCFT on the boundary \cite{ot}.
As we go one step further, the field theory/ M theory duality gives
a supergravity theory on $AdS_4$ or $AdS_7$ for some superconformal theories
in 3 and 6 dimensions, respectively. The maximally supersymmetric theories
in 3 and 6 dimensions have been studied recently 
\cite{aoy,lr,minwalla2,halyo1,gomis}.
The lower supersymmetric case is realized on the worldvolume of M theory
at orbifold singularities \cite{fkpz} (See also 
\cite{berkooz}).
Very recently, along the line of \cite{ot}, 
the Kaluza-Klein states of supergravity theory on the orbifolds of
$AdS_4 \times {\bf S^7}$ were studied and compared with the chiral 
primary operators
of the SCFT on the boundary \cite{eg}. Furthermore, the analysis for 
orbifolds of $AdS_7 \times {\bf S^4}$ was worked out in \cite{aot}.
 Other important developments of the AdS/SCFT duality can be found in
\cite{hor,mald1,mald,ry,go,wi1,li1,coot,keha,bkv}.
  
On the other hand, important results have been obtained by studying the 
aspects of
supersymmetric chiral gauge theories derived from
the Brane Boxes Models \cite{hz,hsu,hu}.
By considering a $D4$ brane stretched between two pairs of two NS5 branes,
the resulting theory is ${\cal N} = 2$ in 3 dimensions.
As in \cite{hu}, we show that a specific Brane Box Model is mapped to a
configuration with D2 branes at ${\bf C^3/Z_3}$ singularity giving
${\cal N} = 2$ in 3 dimensions. Another interesting class of
models consists of web configurations where D3 branes are suspended
between (1, 0), (0, 1) and (1, 1) webs which is again shown to be
mapped to D2 branes at ${\bf C^3/Z_3}$ singularity. We also discuss
possible connection between the Brane Box Model and the web configuration.
 
As a main result of this paper, 
we investigate the Kaluza-Klein states of supergravity theory
on orbifolds of
$AdS_4 \times {\bf S^7}$.
We obtain the chiral primary operators in the superconformal
multiplets by using the correspondence between AdS compactifications and SCFT.
As pointed out in 
\cite{hu} for a four dimensional theory, our method will allow us to obtain 
partial information because
we consider only the untwisted modes which are
accessible in the supergravity approximation which does not give information 
about the twisted sectors. 
 
In section 2 we review some known results for ${\cal N} = 8$ and 
${\cal N} = 4$
in three dimensions.
In section 3 we obtain our results for ${\cal N} = 2$ in three dimensions.
We identify the SCFT and we determine its spectrum of 
chiral primary operators. 
\section{ Review on $AdS_4$ and ${\cal N}=8, 4 $ SCFTs }
\setcounter{equation}{0}

$\bullet$ ${\cal N}=8$ supersymmetric case

Let us consider M theory on $AdS_4 \times {\bf S^7}$ with a 7 form flux of $N$
quanta on ${\bf S^7}$.  The radii of $AdS_4$ and ${\bf S^7}$ are given by
$2R_{AdS_4} = R_{{\bf S^7}} = l_p(32 \pi^2 N)^{1/6}$.  Eleven dimensional
supergravity is appropriate for energies of the order of $1/R_{AdS_4}$ if
$N$ becomes very large. The bosonic symmetries are
given by $SO(3,2) \times SO(8)$.
In \cite{mal} it was proposed that the conformal theory on $N$
parallel M2 branes on the boundary of $AdS_4$ is dual to M theory on
the above background. The $SO(3,2)$ part
of the symmetry of the supergravity side is the conformal group of
the SCFT on the boundary while the $SO(8)$ part
corresponds to the R symmetry of the boundary SCFT.  From the point of
view of type IIA string theory, it is known that 
this SCFT is the strong
coupling limit of the 3 dimensional ${\cal N}=8$ $U(N)$ gauge theory on $N$
coincident D2 branes.

Let us study the correspondence between the Kaluza-Klein
excitations of supergravity and the chiral primary fields of the SCFT.
The spectrum of the
Kaluza-Klein harmonics of eleven dimensional supergravity on $AdS_4
\times {\bf S^7}$ was analyzed in \cite{sugra} some time ago.  
There exist three
families of scalar excitations and two families of pseudoscalar
excitations. Three of them contain states with only positive $m^2$
corresponding to irrelevant operators.  One family
contains states with negative and zero $m^2$ with masses given by
\bea
m^2 =  \frac{1}{4} k (k-6), \;\;\; k=2, 3, \cdots
\label{m1}
\eea
They fall into the $k$th order symmetric traceless
representation of $SO(8)$ with unit multiplicity.
The scaling dimensions of
the corresponding chiral operators \cite{aoy} in the SCFT side
are
\bea
\Delta = \frac{k}{2},  \;\;\; k=2, 3, \cdots
\label{dim1}
\eea
By regarding this  as the strong coupling limit of the 3 dimensional
${\cal N}=8$ SYM theory, some of these operators may be identified with
operators of the form $\mbox{Tr} (X^{i_1} X^{i_2} ... X^{i_k})$, where the
$X^i$ are the scalar fields in the vector multiplet. 
For $k=2, \dots, 5$ these are relevant operators
in SCFT, and for $k=6$ they are marginal.
There is one other family of
pseudoscalar excitations which contains states with
negative and zero $m^2$, corresponding to relevant and marginal operators
in the SCFT.
The masses of this family are given by
$
m^2 = \frac{1}{4} \left ((k-1)(k+1) -8 \right ), \;\;\; k=1, 2, \cdots  
\label{m2}
$
The $k$ th state transforms in a representation of $SO(8)$
corresponding to the product of a ${\bf 35_c}$ with $(k-1)$ ${\bf
8_v}$'s.  
The dimensions of the corresponding
operators in the SCFT are
$
\Delta = \frac{k+3}{2}, \;\;\; k=1, 2, \cdots
\label{dim2}
$
For $k=1$ we have ${\bf 35_c}$ pseudoscalar relevant operators of
dimension 2.  In the UV SYM flowing to this SCFT, we can identify
these operators with a product of two fermions times $k-1$ scalars, as
in \cite{witten,ffz}. 
 
Next we identify one family of vector
bosons that contains massless states. The masses of this family 
are given by
$
{\widetilde m}^2 = \frac{1}{4}(k^2-1), \;\;\; k=1, 2, \cdots
\label{m3}
$
The dimensions of the corresponding
1-form operators in the SCFT are
$
\Delta =\frac{k+3}{2}, \;\;\; k=1, 2, \cdots
\label{dim3}
$
The massless vector at $k=1$ in (\ref{dim3}) corresponds to 
R symmetry current of dimension 2.

$\bullet$ ${\cal N}=4$ supersymmetric case

In \cite{fkpz,gomis} supergravity duals of 3 dimensional ${\cal N}=4$ 
theories with ADE global symmetries were
analyzed.  They can
be realized as worldvolume theories of brane configurations in
spacetime.  The
spacetime  compactification can be  read from the near horizon
geometry of the brane configuration.
The brane configuration corresponding to the fixed point is the theory
on $N$ M2 branes at a ${\bf C^2}/\Gamma$ singularity.  The near
horizon geometry of this  brane configuration  is $AdS_4\times
{\bf S^3}\times_f{\bf D_4}/\Gamma$ 
where the ${\bf S^3}$ is fibered over the $\Gamma$
quotiented four disk. The theory dual to the fixed point is
M theory on $AdS_4\times {\bf S^3}\times_f{\bf D_4}/\Gamma$ \cite{gomis}. 

Only a subset of the allowed projected states will
have the right quantum numbers to be in short ${\cal N}=4$ supersymmetry
multiplets. The superconformal primaries from the scalar family are
given by in terms of $SU(2)_{ADE} \times SU(2)_L \times SU(2)_R$ 
representation.
None of the surviving states from the pseudoscalar tower have the
right  quantum numbers to be ${\cal N}=4$ SCFT. 
The massless vector and the graviton correspond  to
conserved currents in the superconformal field theory. They couple to
the global symmetry current in the adjoint of the R symmetry and to
the energy momentum respectively. 

\section{ An Orbifold of $AdS_4 \times {\bf S^7}$ and ${\cal N}=2$ SCFT  }
\setcounter{equation}{0}
\subsection{Singularities and Brane Configurations}
 We now take orbifolds\footnote{
Generally speaking, there are a variety of orbifolds with free or nonfree
actions on ${\bf S}^7$ leading to different amount of supersymmetry. 
See recent 
paper by Morrison and Plesser\cite{mp}. Let us consider
M2 branes at ${\bf C}^4/\Gamma$ singularity and
that the group $\Gamma$ is generated by $\mbox{diag}( 
 \mbox{exp}(2\pi i/k), \mbox{exp} (-2 \pi i/k), \mbox{exp}( 2 \pi i a/k),
\mbox{exp} (-2\pi i a/k))$ for some relatively prime intergers $a$ and $k$.
If $a=1, k=2$ we get maximal case ${\cal N}=8$.
For $a=\pm 1, k \geq 2$, one gets ${\cal N}=6$ 
theory\cite{halyo2} where the corresponding field theory duals are present.
When $\Gamma$ is a binary dihedral group, $D$ type, singularity and we embed
$\Gamma$ into $SU(2) \times SU(2)$, we get ${\cal N}=5$ theory\cite{dn}.
When $a \neq \pm 1$, the theory has ${\cal N}=4$ supersymmetry. It is not 
clear at the moment 
how ${\cal N}=3, 1$ cases are realized by some orbifolds. }
of $AdS_4 \times {\bf S^7}$. As usual, we want
an orbifold action only on ${\bf S^7}$. We want to see the near horizon 
geometry of the M2 branes. 
We use orbifold construction which gives ${\cal N} = 2$
SCFT in 3 dimensions but the transverse space is not compact. 
To do this we take a brane configuration
with M2 branes at ${\bf C^3}/\Gamma$ singularity\footnote{
The group actions $\Gamma $ contains A type( cyclic group ), D type(
binary dihedral group, E type( binary tetrahedral, octahedral, icosahedral 
group ) in general. For the AdS/CFT correspondence, in the field 
theory side we are considering, the global symmetry is given by $U(1)_R$.
In the supergravity side this corresponds to isometry of ${\bf S^7}$. 
Somehow $SO(8)$ breaks into $U(1)_R$ times other part. We found that the only
${\bf Z_3}$ orbifold case in cyclic group is relevant to this analysis.
Other orbifold cases are not too much interesting in this sense. Of course, 
they will give rise to different global symmetry.}.
By the connection of the
M theory configuration involving M2 branes and a type IIA vacuum, 
we can think of type IIA vacuum as a compactification of M theory on
${\bf R^{1, 9} \times S^{1}}$. Then the D2 branes correspond to
M2 branes unwrapped around $x^{10}$. 
 In 11 dimensions
the transverse space to the M2 brane is topologically ${\bf R^8}$ and we act 
with $\bf Z_3$ to form ${\bf R^8/Z_3 = R^2 \times C^3/Z_3  }$. 
Here ${ \bf Z_3}$ acts
on ${\bf C}^3$ by $\zeta \cdot (z_1, z_2, z_3) = (\omega z_1, 
\omega z_2, \omega z_3)$
where $\omega = \exp (2\pi i/3)$ and $\zeta$ is the generator of ${\bf Z}_3$.
We first reduce the theory from 11 dimensions to 10 
dimensions (type IIA string theory) to obtain 
D2 branes with transverse space $ {\bf R} \times {\bf C^3/Z_3}$ and we can
easily determine the global symmetry.

{ \it If we take $x^3$ to be compact we obtain
D2 branes in ${\bf S^1 \times C^3/Z_3}$.
The isometry of ${\bf S^1}$ is a global symmetry
for the field theory on the world volume of the D2 branes and is
identified with the $U(1)_R$ symmetry. We then have the
global symmetry breaking $SO(8) \rightarrow SU(3) \times U(1)_R$
because ${\bf C^3/Z_3}$ is 
a Calabi-Yau threefold having $SU(3)$ holonomy group }.

Here we have two identifications between isometries of the compactified
space and symmetries on the boundary. Firstly we have the Killing
spinor equation in the spacetime metric which gives a condition for the
unbroken supersymmetries in the spacetime and for the case of 
${\bf S^1 \times C^3/Z_3}$ this gives eight unbroken supersymmetries 
corresponding to ${\it N} = 2$ superconformal field theories in 3
dimensions (we use the  fact that compactification on ${\bf C^3/Z_3}$
breaks 1/4 of the supersymmetry and compactification on ${\bf S^1}$ does
not break any more supersymmetry. Secondly, the $R$ symmetry comes from the
isometry of ${\bf S^1}$ which gives a $U(1)$ symmetry which is just the 
required $R$ symmetry for ${\it N} = 2$ in 3 dimensions. 

We also need to explain the passage from the supegravity solution on
$AdS_4 \times {\bf S^7}$ to a solution on ${\bf S^1 \times C^3/Z_3}$.
Although the eleven dimensional solution by promoting D2 brane
solution is not exactly M2 brane solution, in general, when the eleventh
direction is compact, by taking M2 branes to be localized in the transverse
directions, these two solutions are the same in M theory limit\cite{mald1}.
The passage from D2 to M2 is by dimensional reduction, both being extended
in the $(x^1, x^2)$ directions. By using the fact that the
radius of ${\bf S^7}$ is proportional with $N$ and is thus very large in the
large $N$ limit, the $x^3$ direction can be approximated with a
circle in the large $N$ limit.
   
We can now identify the gauge symmetry and the field content.
The starting point is the configuration with D2 branes
in ${\bf S^1 \times C^3/Z_3}$, the fact that the D2 branes are at a
singularity implying 
that the gauge group of the
theory is $SU(N) \times SU(N) \times SU(N)$ and 
the matter content is given by fields
transforming as ${\bf (N, \overline{N}, 1) 
\oplus (1, N, \overline{N}) \oplus (\overline{N}, 1, N)}$. 

We make here a connection with two related brane configurations.
We firstly use the result of \cite{hu} where they discussed the
${\cal N} = 1$ theory in 4 dimensions. One of the main results of their work
is that a configuration with D3 branes at ${\bf C^3/Z_3}$ singularities
is mapped to a Brane Box Model of intersecting NS5 and D5 branes 
constructed on a two-torus $T^2$. The map is interpreted as a T-duality
along the two directions of the torus. 
Here we are interested in 3 dimensional
${\cal N}= 2$ theories so we take a Brane Box Model with intersecting 
NS5 and D4 branes which can be obtained from the configuration of \cite{hu} by 
a T-duality along a compact direction parallel with the NS5 branes
which leaves unchanged the two NS5 branes.  

For self-consistency, we repeat their arguments here. Let us consider
the type IIA theory obtained from M theory after dimensional reduction with:
NS5 branes along $(x^0, x^1, x^2, x^3, x^4, x^5)$ directions,
NS'5 branes along $(x^0, x^1, x^2, x^3, x^6, x^7)$ directions
and D4 branes along $(x^0, x^1, x^2, x^4, x^6)$ directions

The D4 branes are finite in the 4 and 6 directions so their low-energy
effective world volume theory is $2 + 1$ dimensional and the supersymmetry 
is $1/8$ of the original supersymmetry (so we are dealing with $N = 2$ 
supersymmetry in 3 dimensions). We take $x^4$ and $x^6$ directions to be 
circles
so we can make T-duality on them. We divide the $x^4$ and $x^6$ 
plane into a set 
of boxes, the number of boxes on $x^4$ and $x^6$ 
directions giving the gauge group
and the matter content, respectively. 
We are interested here in the brane box of figure 7, in particular,
in \cite{hu}. This is a $3 \times 1$ configuration in the $x^4$ and $x^6$ 
directions
having the right gauge group and matter content. 
To study explicitly the connection between the Brane Box 
configuration and the one involving 
D2 branes at singularity we need to perform a T 
duality along $x^4$ and $x^6$ directions so the
D4 branes are mapped into D2 branes along $(x^0, x^1, x^2)$ 
on top of ${\bf C^3/Z_3}$ as explained 
in page 29 of \cite{hu}. 

Another brane configuration equivalent to M2 branes with a transverse space
${\bf R^2} \times {\bf C^3/Z_3}$ is obtained by using the duality
between 
M theory compactified on $T^2$ and
type IIB theory compactified on $S^1$. As in \cite{ah,k,ahk}, 
M2 branes unwrapped on 
$T^2$ correspond to D3 branes wrapped on $S^1$ and the M5 branes 
wrapped on $(p, q)$ cycles of $T^2$ correspond to $(p, q)$ 
branes in type IIB theory and both correspond to a 2 brane in
9 dimensions. 
The space ${\bf {\bf C^3/Z_3}}$ can be considered as an affine cone over
a projective space ${\bf P}^2$. By blowing up the vertex, the space can
be identified with a neighborhood of ${\bf P}^2$ embedded in Calabi-Yau
threefold, in which the normal bundle of ${\bf P}^2$ in the Calabi-Yau
threefold is the canonical line bundle. We now have a 3-dimensional
local toric geometry, where the extra circle action comes from the rotation
on the phase of the normal line bundle. Now by blowing down
${\bf P}^2$, we obtain an affine toric variety ${\bf {\bf C^3/Z_3}}$
whose toric skeleton is given by a vertex with 3 external legs \cite{lv}.
So the $(1, 0), (0, 1), (1, 1)$ brane web configuration corresponds
to  ${\bf C^3/Z_3}$ toric skeleton. 
If we have N D3 branes between two vertices, the gauge theory is
$SU(N)$ with a chiral field transforming in the  
$N$ representation. 

Now, if we return to the Brane Box Model of above, we can connect it to
the $3 \times 1$ model by a set of dualities
we can arrive to the right web configuration which contains
three web points and D3 branes between them on a circle giving thus
the right gauge group and field content as explained in \cite{ah}.
To see this in detail, we start in 11 dimensions from a configuration
with a M2 brane in (012) directions and two Kaluza - Klein monopoles
in (589,10) and (6789) directions. By reducing to 10 dimensions
to obtain type IIA theory, we obtain a configuration with a
D2 brane in (012) directions, a D6 brane in (0123467) directions and a
KK monopole in (6789) directions. A T-duality on $x^6$ direction gives
D3(0126), NS(012345) and D5(012347). As explained in \cite{ah}, in
the $(x^5, x^7)$ plane besides the (1,0) and (0,1) lines represented
by NS and D5 there is the (1,1) line so the configuration gives just
a D3 brane between two webs. On the other side, if we start in 11 dimensions
with a configuration with M2 (012), KK (6789) and KK (4589),  
by reduction to type IIA and two T-dualities on $x^4, x^6$ directions, one
obtains a configuration with D4(1246), NS (12345) and NS (12367) which is
just our Brane Box Model. In 11 dimensions the two starting brane
configurations are almost identical because all the dimensions
are infinite, therefore quantum results concerning Brane Box Models
could be obtained by using similar results for web models \cite{k,ahk}.
\subsection{Surviving Kaluza-Klein modes}
We now proceed to identify the Kaluza-Klein modes that survive after 
the orbifolding procedure. We also determine the operators 
in the conformal field theory side to
which the Kaluza-Klein modes couple. These operators are built as combinations
of the scalars which enter the theory. 
These are of three types: the 
scalar part of the chiral multiplet coming from the 4 dimensional
chiral multiplet, the real scalar corresponding to the component of the
$D = 4$ vector potential in the reduced direction and the third one is
obtained in the bulk of the Coulomb branch by dualising the gauge fields.
The last two are combined into a chiral superfield $\Phi^j, j = 1, 2, 3$. 
We denote the
matter multiplets by $U, V, W$. We assign the $U(1)_{R}$ charges
1/2 to $\Phi^j$ and 2/3 to $U, V, W$. By considering the
relation between the dimension of chiral operators and $U(1)_{R}$
symmetry charges as \cite{chk,sei,minwalla1}:
\beq
\Delta = |R| \;\;\; \mbox{or} \;\;\; |R| + 1,
\eeq 
we take the dimension of $\Phi^j$ as 1/2 and the one of $U, V, W$ as 2/3.
Now we proceed to identify the surviving Kaluza-Klein modes.  

$\bullet$ 
The $k=2$ Kaluza-Klein particle in (\ref{m1}) transforms in the
$ {\bf 35_v} $ of $SO(8)$. By decomposing $ {\bf 35_v} $ into the 
representation
of $SU(3)_{\Gamma} \times U(1)_R$, we obtain 
${\bf 1_0}$ which is invariant under $\Gamma $ with right $U(1)_R$ quantum
number.
We list $SO(8) \rightarrow SU(3)_{\Gamma} \times U(1)_R$ branching rules
in the Appendix with the help of \cite{ps}.
We expect that a dimension 1 chiral primary operator,
according to (\ref{dim1}),
 to live in the 
boundary SCFT.
This Kaluza-Klein mode couples to the dimension 1 chiral operator 
$\Sigma_{i=1}^{3} \mbox{Tr} (\Phi^{i} \widetilde{\Phi}_{i})$ 
where the index $i$ enumerates the 
three gauge groups $SU(N) \times SU(N) \times SU(N) $. 
Here $\widetilde{\Phi}_{i}$ is the field conjugate to 
$\Phi$ and has $U(1)_{R}$ charge $-1/2$.

$\bullet$ The The $k=3$ Kaluza-Klein particle in (\ref{m1}) transforms in the
$ {\bf 112_v} $ of $SO(8)$.
There is no invariant  state under the 
$\Gamma$ projection with right $U(1)_R$ charge.

$\bullet$ The $k=4$ Kaluza-Klein particle in (\ref{m1}) transforms in the
$ {\bf 294_v} $ of $SO(8)$.
The ${\bf 1_2, 1_{-2}, 8_2, 8_{-2}, 10_{-2}}$ and ${\bf {\overline{10}}_2 }$ 
are invariant under $\Gamma$ so these states
will survive the projection. From their quantum numbers, 
this Kaluza-Klein modes couple respectively to
$\mbox{Tr} (U^{i} V^{j} W^{k})$ and $\mbox{Tr} (\widetilde{U}^{i} 
\widetilde{V}^{j}
\widetilde{W}^{k})$
where here we take all the possible
combinations of $i, j$ and $k$ by using the product rules 
${\bf 3 \times 3 = \overline{3} \oplus 6, 3 \times \overline{3} = 1 \oplus 8,
3 \times 6 = 8 \oplus 10}$. 
So all the surviving Kaluza-Klein modes can couple to
chiral primary operators in the SCFT.

$\bullet$ The $k=5$ Kaluza-Klein particle in (\ref{m1}) transforms in the
$ {\bf 672'_v} $ of $SO(8)$.
There is no invariant  state under the $\Gamma$ projection.

$\bullet$ The $k=6$ Kaluza-Klein particle in (\ref{m1}) transforms in the
$ {\bf 1386_v} $ of $SO(8)$.
The $
{\bf 1_2, 1_{-2}, 8_2, 8_{-2}, 10_{-2}, 10_2, 
{\overline{10}}_2,  {\overline{10}}_{-2},
27_2, 27_{-2}, 28_{-2}, {\overline{28}}_{2}, 35_{-2}  } 
$ and ${\bf {\overline{35}}_2}$
are invariant under $\Gamma$ so these states will survive the projection.
They have right $U(1)_R$ charge.
These Kaluza-Klein modes can couple to dimension 3 operators which are
similar to the ones obtained at $k = 4$, being as
$\mbox{Tr} (U^{i} V^{j} W^{k} \Phi^{i} \widetilde{\Phi}_{i})$.
The last six Kaluza - Klein modes cannot be written in terms of
the short distance fields. The only gauge invariant possibility would
involve $\mbox{Tr} (U^{i} V^{j} W^{k})^{2}$ which is of dimensions 4
so being greater than 3.

If we look now to the pseudoscalar tower, we see that for $k = 1$
( $\Delta = 2$ ) we have ${\bf 1_2, 1_{-2}}$ as surviving states
and for $k = 3$ we have the ${\bf 1_2, 1_{-2}, 8_{2}, 8_{-2}}$
as surviving states, all of them can be written as gauge invariant combinations
of operators in SCFT.
The massless vector and the graviton couple to the global $U(1)_{R}$
symmetry current and to the energy momentum respectively, their dimensions
being protected from quantum corrections.

\section{ Conclusion }
\setcounter{equation}{0}

In this paper we have obtained a part of the spectrum of chiral primary 
operators 
in the superconformal
multiplets of ${\cal N}= 2, D = 3$ SCFT by using the 
field theory on the M2 worldvolume.
We performed only calculations in supergravity so we have just obtained the
untwisted sector results. In order to smoothen out the singularity of the
spacetime, we need to consider the full M theory. 
It would be interesting to study the possible twisted modes in the full 
M theory.

We have used the map between M2 branes at ${\bf C^3/Z_3}$ and Brane Box Models.
Another configuration giving M2 branes at ${\bf C^3/Z_3}$ is 
the one with D3 branes between three webs of (0,1), (1, 0) and (1, 1) branes
which are on a circle.
We have identified a connection between a Brane Box Model and a web model
which could help to enlarge the quantum description of both.  

Our approach to obtain the spectrum of chiral primary operators 
is similar to the one used in 
\cite{kachru} for D3 branes at ${\bf C^3/Z_3}$ which gives
${\cal N} = 1$ in $D = 4$. 
It would be interesting to obtain ${\cal N} = 2$ in $D = 3$ by considering 
M theory compactified on a Calabi-Yau 4-fold. One discussion is made in
\cite{lv} where a Calabi-Yau 4-fold is described in terms of a
tetrahedron corresponding to certain $(p, q, r)$ 4 - branes where
the $(p, q, r)$ vector label the Kaluza - Klein monopoles obtained when
M theory is compactified on $T^{3}$.

\vspace{2cm}

\centerline{\Large \bf Acknowledgments} 

We would like to thank J. Gomis, A. Hanany, S. Kachru, B. Kol, J. Maldacena,
S. Minwalla, C. Vafa for correspondence, discussions and important
comments on the manuscript. 
Kyungho Oh is supported in part by UM Research Board and APCTP.
This work of Changhyun Ahn 
was supported (in part) by the Korea Science and Engineering 
Foundation(KOSEF) through the Center for Theoretical Physics(CTP) at Seoul 
National University. 

We would like to thank our referee for important comments on a previous
version of this manuscript.
\section{Appendix: $SO(8)$ Branching Rule }

\begin{tabular}[b]{|c|l|l|}
\hline Fields  & $SO(8)$ Dynkin label &  $SU(3)_{\Gamma} \times U(1)_R$ \\ 
\hline
vector
& $(0, 1, 0, 0): \;\;\; {\bf 28}$    & 
${\bf 1_{0, 0} \oplus 3_{2/3, 2/3, -4/3} \oplus 
{\overline{3}}_{-2/3, -2/3, 4/3} \oplus
8_0} $   \\ \hline
scalar  & $(2, 0, 0, 0): \;\;\; {\bf 35_v}$  
              & ${\bf  1_{0, -2, 2} \oplus 3_{2/3, -4/3} \oplus 
{\overline{3}}_{-2/3, 4/3} } $ \\
& & $ {\bf  \oplus 6_{-2/3}
\oplus {\overline{6}}_{2/3} \oplus 8_0}  $   
\\ \hline
pseudoscalar 
& $(0, 0, 2, 0): \;\;\; {\bf 35_c}$ & 
 ${\bf  1_{0, -2, 2} \oplus 3_{2/3, -4/3} \oplus 
{\overline{3}}_{-2/3, 4/3} } $ \\
& & $ {\bf  \oplus 6_{-2/3}
\oplus {\overline{6}}_{2/3} \oplus 8_0}  $   
 \\ \hline 
scalar & $(3, 0, 0, 0): \;\;\; {\bf 112_v}$    & 
${\bf   1_{-1, 1, -3, 3} \oplus 
 3_{-1/3, 5/3, -7/3}  \oplus {\overline{3}}_{1/3, -5/3, 7/3} } $ \\ 
& & ${\bf \oplus 6_{1/3, -5/3}   \oplus  {\overline{6}}_{-1/3, 5/3}  
\oplus 8_{-1, 1}  \oplus
10_{-1}  } $ \\
& & $ {\bf \oplus {\overline{10}}_1 
\oplus 15_{-1/3} \oplus {\overline{15}}_{1/3}} $
\\ \hline
pseudoscalar & $(1, 0, 2, 0): \;\;\; {\bf 224_{cv}}$  
              & ${\bf 1_{-1, -1, 1, 1, -3, 3} } $ \\
& & $ {\bf  \oplus 
3_{-1/3, -1/3, -1/3, 5/3, 5/3, -7/3, -7/3}  }$ \\
& & $ {\bf  \oplus {\overline{3}}_{1/3, 1/3, 1/3, -5/3, -5/3, 
7/3, 7/3}  } $ \\ 
& & $ {\bf \oplus 
6_{1/3, 1/3, -5/3, -5/3} 
\oplus  {\overline{6}}_{-1/3, -1/3, 5/3, 5/3}} $ \\
& & $ {\bf  \oplus 8_{-1, -1, -1, 1, 1, 1}  \oplus 10_{-1} \oplus 
{\overline{10}}_1 } $ \\
& & $ {\bf \oplus 15_{-1/3, -1/3} \oplus {\overline{15}}_{1/3, 1/3} }$ \\
\hline
scalar   
& $(4, 0, 0, 0): \;\;\; {\bf 294_v}$ 
& ${\bf 1_{0, -2, 2, -4, 4}  \oplus
 3_{2/3, -4/3, 8/3, -10/3} } $ \\
& & $ {\bf \oplus {\overline{3}_{-2/3, 4/3, -8/3, 10/3}}  
\oplus 6_{-2/3, 4/3, -8/3} } $ \\
& & $ {\bf  \oplus 
{\overline{6}}_{2/3, -4/3, 8/3} \oplus 8_{0, -2, 2}  \oplus 10_{0,-2} \oplus
{\overline{10}}_{0, 2} } $ \\ 
& & $ {\bf \oplus 15_{2/3, -4/3} \oplus 15'_{-4/3} \oplus 
{\overline{15'}}_{4/3}  } $ \\
& & $ { \bf \oplus {\overline{15}}_{-2/3, 4/3} 
\oplus 24_{2/3} \oplus {\overline{24}}_{-2/3} \oplus 27_0}$ \\ \hline
scalar & 
$(5, 0, 0, 0): \;\;\; {\bf 672'_v}$ & 
${\bf 1_{-1, 1, -3, 3, -5, 5} \oplus  3_{-1/3, 5/3, -7/3, 11/3, -13/3} } $ \\
& & $ {\bf  \oplus {\overline{3}}_{1/3, -5/3, 7/3, -11/3,13/3} 
\oplus   6_{1/3, -5/3, 7/3, -11/3}   } $ \\
& & ${\bf  \oplus {\overline{6}}_{-1/3, 5/3, -7/3, 11/3}
\oplus 8_{-1, 1, -3, 3} \oplus 10_{-1, 1, -3} }$ \\
& & ${\bf \oplus {\overline{10}}_{-1, 1, 3}  
\oplus 15_{-1/3, 5/3, -7/3} \oplus 
{\overline{15}}_{1/3, -5/3, 7/3} } $ \\
& & $ {\bf 
\oplus 15'_{-1/3, -7/3} \oplus {\overline{15'}}_{1/3, 7/3}  }$ \\
& & ${\bf \oplus {\overline{21}}_{-5/3} \oplus 21_{5/3} 
\oplus 24_{-1/3, 5/3} \oplus
 {\overline{24}}_{1/3, -5/3} }$ \\
& & ${\bf \oplus 35_{-1} \oplus {\overline{35}}_{1} \oplus 27_{-1, 1}
\oplus {\overline{42}}_{1/3} \oplus 42_{-1/3}}$ \\ \hline
\end{tabular}

\begin{tabular}[b]{|c|l|l|}
\hline Fields  & $SO(8)$ Dynkin label &  $SU(3)_{\Gamma} \times U(1)_R$ \\ 
\hline
pseudoscalar & $(2, 0, 2, 0): \;\;\; {\bf 840'_s}$ & 
${\bf  1_{0, 0, 0, -2,-2, 2, 2, -4, 4}  } $ \\
& & $ {\bf \oplus 3_{2/3, 2/3, 2/3, 2/3, -4/3, -4/3, -4/3, -4/3} } $ \\  
& & $ { \bf \oplus 3_{8/3, 8/3,  -10/3, -10/3}  }$ \\ 
&& $ {\bf \oplus {\overline{3}}_{-2/3, -2/3, -2/3, -2/3, 4/3, 
4/3, 4/3, 4/3 }} $ \\
& & $ { \bf \oplus {\overline{3}}_{ -8/3, -8/3, 10/3,
10/3} }$\\ 
&& $ {\bf \oplus  6_{-2/3, -2/3, -2/3, -2/3, 4/3, 4/3, 
4/3, -8/3, -8/3, -8/3}   }$\\ 
&& $ {\bf \oplus   {\overline{6}}_{2/3, 2/3, 2/3, 2/3, -4/3, -4/3, 
-4/3, 8/3, 8/3, 8/3} }$\\ 
&& $ {\bf  \oplus 8_{0, 0, 0, 0, 0, 0, -2, -2, -2, -2, 2, 2, 2, 2 } }$\\ 
&& $ {\bf \oplus 10_{0, 0, -2, -2} \oplus  {\overline{10}}_
{0, 0, 2, 2}} $ \\ 
&& $ {\bf \oplus 15_{2/3, 2/3, 2/3, 2/3, -4/3, -4/3, -4/3, -4/3} } $ \\
& & $ {\bf  \oplus {\overline{15}}_{-2/3, -2/3, -2/3, -2/3, 4/3, 
4/3, 4/3, 4/3} } $ \\
& & $ {\bf   \oplus 15'_{-4/3} \oplus {\overline{15'}}_{4/3}   
\oplus {\overline{24}}_{-2/3, -2/3}  \oplus 24_{2/3, 2/3} } $ \\
& & $ {\bf \oplus 27_{0, 0, 0} }$ \\  
\hline
scalar 
& $(6, 0, 0, 0): \;\;\; {\bf 1386_v} $ & ${\bf 1_{0, -2, 2, -4, 
4, -6, 6} } $ \\
& & $ { \bf \oplus  3_{2/3, -4/3, 8/3, -10/3, 14/3, -16/3} } $ \\
& & $ {\bf  \oplus {\overline{3}}_{-2/3, 4/3, -8/3, 10/3, -14/3, 16/3}  } $ \\
& & $ {\bf \oplus  6_{-2/3, 4/3, -8/3, 10/3, -14/3} } $ \\    
& & $ {\bf \oplus  {\overline{6}}_{2/3, -4/3, 8/3, -10/3, 14/3}  }$ \\
&& $ {\bf \oplus 
8_{0, -2, 2, -4, 4} \oplus 10_{0, -2, 2, -4} \oplus 
{\overline{10}}_{0, -2, 2, 4}  }$ \\
&& $ {\bf  \oplus
15_{2/3, -4/3, 8/3, -10/3}  \oplus 15'_{2/3, -4/3, -10/3} } $ \\ 
& & $ {\bf  \oplus {\overline{15}}_{-2/3, 4/3, -8/3, 10/3} 
\oplus {\overline{15'}}_{-2/3, 4/3, 10/3} } $ \\
& & $ {\bf  \oplus 21_{2/3, 8/3} 
\oplus {\overline{21}}_{-2/3, -8/3} \oplus24_{2/3, -4/3, 8/3} } $ \\
& & $ { \bf \oplus {\overline{24}}_{-2/3, 4/3, -8/3} 
\oplus 27_{0, -2, 2}  \oplus 28_{-2} \oplus {\overline{28}}_{2} }$ \\
&& $ {\bf \oplus 35_{0, -2} \oplus {\overline{35}}_{ 0, 2} 
\oplus 42_{2/3, -4/3} \oplus {\overline{42}}_{-2/3, 4/3} }$ \\
&& $ {\bf  \oplus 48_{-4/3} \oplus
{\overline{48}}_{4/3} \oplus 60_{2/3} \oplus {\overline{60}}_{-2/3} 
\oplus 64_0 }$  \\
\hline
\end{tabular}

Table 1. The superconformal multiplets, their $SO(8)$ Dynkin labels and
branching rules for $SU(3)_{\Gamma} \times U(1)_R$. $U(1)_R$ charges are
given in the subscript of $SU(3)_\Gamma$ representations. For simplicity,
we denote ${\bf 1_0 \oplus 1_0}$ by ${ \bf 1_{0, 0}}$ and so on.


\end{document}